# Study on Sub monolayer Epitaxy Growth under Anisotropic Detachment

J. Devkota[1] and S.P. Shrestha[2]
[1]*Central Department of Physics, Tribhuvan University, Kirtipur, Nepal*
[2]*Department of Physics, Patan Multiple Campus, Tribhuvan University, Patan Dhoka, Nepal*

## Abstract

We have performed Kinetic Monte Carlo simulation to study the effect of diffusion anisotropy and bonding anisotropy on island formation at different temperatures during the sub-monolayer film growth in Molecular Beam Epitaxy. We use simple cubic solid on solid model and event based Bortz, Kalos and Labowitch (BKL) algorithm on Kinetic Monte Carlo method to simulate the physical phenomena. We have found that surface anisotropy has no significant role on island elongation however it influences on the island morphology, growth exponent and island size distribution. Elongated islands were obtained when bonding anisotropy was included.

## 1. Introduction

Much of recent interest in the studies of metallic and semiconductor systems has been devoted to epitaxially grown nanostructures not only due to their immense potential for technological applications [1, 2, 3], but also because different surfaces exhibit a surprisingly rich variety of theoretically interesting phenomena [4]. Electronic devices, coatings, displays, sensors and numerous other technologies all depend on the quality of the deposited thin films. Therefore, thin film structures and epitaxially grown sub monolayers with smooth interfaces are of fundamental importance in the field of modern technologies related to electronics, optics and magnetism.

Molecular beam epitaxy (MBE) has become a powerful tool to create model system for this fastest growing research field of the surface science. The fundamental physical processes in MBE involve nucleation by two or more atoms, aggregation of adatoms on the island edges and coalescences of two or more islands. These processes lead to the formation and distribution of islands with various sizes. Understanding the factors governing the shape of clusters in diffusion-controlled aggregation is very essential in the MBE thin film growth.

These days computational methods have become a vital tool for investigating the properties and understanding the theories of metal and semiconductor surfaces [4]. The computer simulation has become an attractive method for studying the surface phenomena. By modeling the key steps in epitaxy and describing the behavior of individual atoms, a simulation can be constructed and tracked with time [5].

We perform kinetic Monte Carlo (KMC) computer simulation in modeling homoepitaxial growth on different types of surfaces using solid-on-solid model and BKL algorithm. It deals with the method of solving the kinetic equations [4]. We consider simple a cubic lattice structure, the simplest type of surface structure for modeling various physical processes. Since the multilayer surface morphology is basically dependent on the surface morphology and size distribution in submonolayer [6], we restrict our attentions in the submonolayer growth. Several groups have studied the evolution of growth of island morphologies as well as scaling of islands during submonolayer epitaxy using Monte Carlo simulation [7-13]. Using various models, effect of anisotropy of diffusion [14-17] and binding strength [12, 17-19] between adatoms and island on island shape were studied as well. The influence of diffusion anisotropy and binding strength between adatoms and island border on island aspect ratio have been studied using temperature, surface diffusion barrier and adatom binding energy as parameter [19]. Using solid on solid model of the epitaxial films growth based on random deposition the influence of critical number of adatoms lateral bonds and the deposition rate on possible surface morphology anisotropy were also studied [20]. In this work, we present the role of diffusion anisotropy and binding anisotropy on island morphology, aspect ratio and growth exponent.

**Growth Model and KMC Simulations**

A Kinetic Monte Carlo simulation method is a probabilistic simulation method that deals with kinetic processes. Each configurational change corresponds to the real events and each event can happen with some probability that depends on their rate. Our KMC simulation model resembles the solid on solid model on a square lattice as used by Heyn [19]. Atoms are randomly deposited on a site of 300x300 square lattices with the flux F = 0.1ML/s at a definite temperature to achieve desired submonolayer coverage θ = 0.075ML. Each deposited atom is then allowed to hop to adjacent sites on the substrate with rate given by

$$D_j = D_o \exp(-E_j/k_B T) \qquad (1)$$

Where,
j = X, Y corresponds to the diffusion direction,
$D_o \approx 10^{13}$ $sec^{-1}$ is common prefactor taken as vibration frequency,
$k_B$ = Boltzmann constant
T = the substrate temperature and
$E_j = E_{S,j} + n_j E_{N,j}$ is the energy barrier for a given process along respective direction.
Here $E_{S,j}$ is a site-independent surface term and $n_j E_{N,j}$ is the term given by the number of occupied lateral nearest neighbors $n_j$ = 0, 1, 2 along j direction with in-plane bond energy $E_{N,j}$. Thus in order to work with both the isotropic and anisotropic crystal surfaces with two non equivalent directions X and Y in our model, the energy terms are written in two fold manner. With these considerations, the energy barrier for the surface diffusion along the X-direction and Y-direction respectively become,

$$E_X = E_{S,X} + n_X E_{N,X} + n_Y E_{N,Y} \qquad (2)$$

$$E_Y = E_{S,Y} + n_X E_{N,X} + n_Y E_{N,Y} \qquad (3)$$

In order to study the influence of terrace diffusion and detachment separately on island elongation, we have monitored the island morphology and aspect ratio as a function of growth temperature over a large range 560K to 1250K in two different series of simulations. In first series only the surface diffusion of monomers is considered. When an adatom encounters another adatom as its nearest neighbor, both atoms are frozen and a stable nucleus of two atoms is formed on the surface. Similarly, when a diffusing adatom encounters an existing island of size S, it sticks on the island yielding an island of size S+1. The dimers and larger islands are assumed to be immobile.

Adatoms on an island edge with only one bond are allowed to detach from the island in second series of our work. Edge diffusion is forbidden because binding energy very low. The rate of detachment is estimated by taking independent energy barrier $E_{diss,j}$. The transition barriers for the various atomistic processes used are similar to the barriers used by Heyn [19]. It should be noted that the absolute values of temperatures are not important in the simulation since temperature scales with energy parameter. Simulation is carried out until the desired coverage is achieved and the quantities of interest are calculated.

**Results and Discussion**

In the first series of simulations, we study the influence of anisotropic terrace diffusion using the model for irreversible island growth. The neighbor related energy terms are set to such a high value ($E_{N,X} = E_{N,Y}$ = 10 eV) that detachment and edge diffusion is suppressed. The cross-channel (along Y- axis) diffusion energy barrier is taken constant ($E_{S,Y}$ = 1.3 eV) through out the simulation. The degree of anisotropy is varied by setting the different values of in-channel (along X- axis) substrate diffusion energy barrier $E_{S,X}$=1.3 eV, 1 eV, and 0.8 eV.

Fig.1 (a-f) depicts the island morphologies for three different values of in-channel diffusion barrier $E_{S,X}$ =1.3 eV, 1 eV and 0.8 eV at substrate temperatures T= 715K and 835K respectively. The horizontal panel from left to right depicts the variation of the island morphologies with in-channel diffusion barrier for two different fixed values of the temperatures (first row is for T=715K and second row is for T=835K). The study of variation of island morphology with the variation in in-channel diffusion barrier at temperature T= 715K [Fig.1 (a-c)] shows that at low diffusion barrier, the values of the island number density is low but the size of island is large. Similar trend was observed in the island morphologies for other higher temperatures.

Fig.1 (a,d,g,j) depicts the variation of island morphology with temperature for different fixed value of $E_{sx}$ in the vertical panels. From the figure, we observe that island number density decreases with increase in the temperature where as the size of island increases. The island morphologies for other values of diffusion barrier

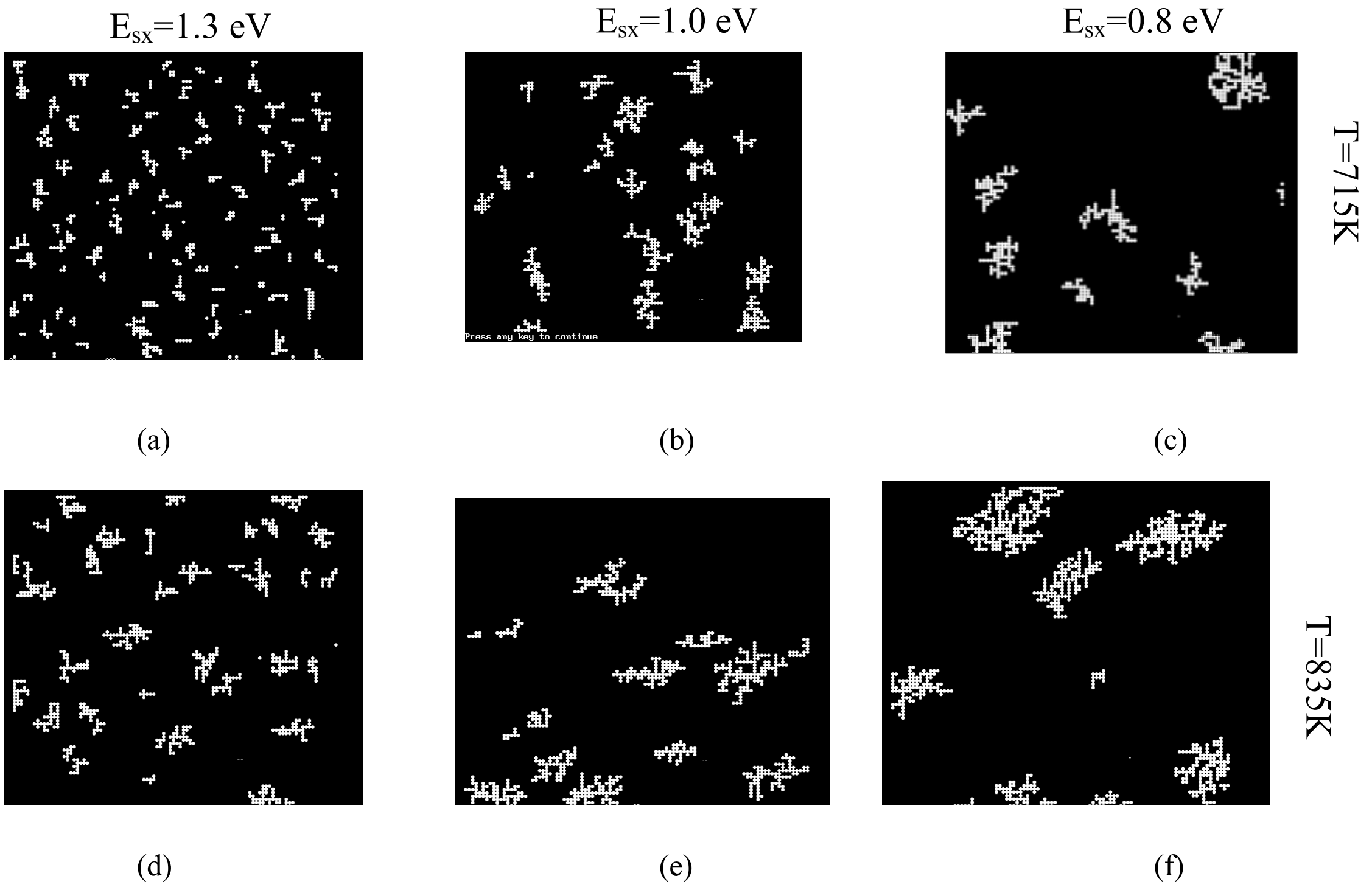

Fig.1. Section 140*140 of the typical island morphology versus temperature for different values of substrate diffusion energy for model – I with $E_{sy}$ = 1.3 eV, flux F = 0.1 ML/s and cov. θ = 0.075ML.

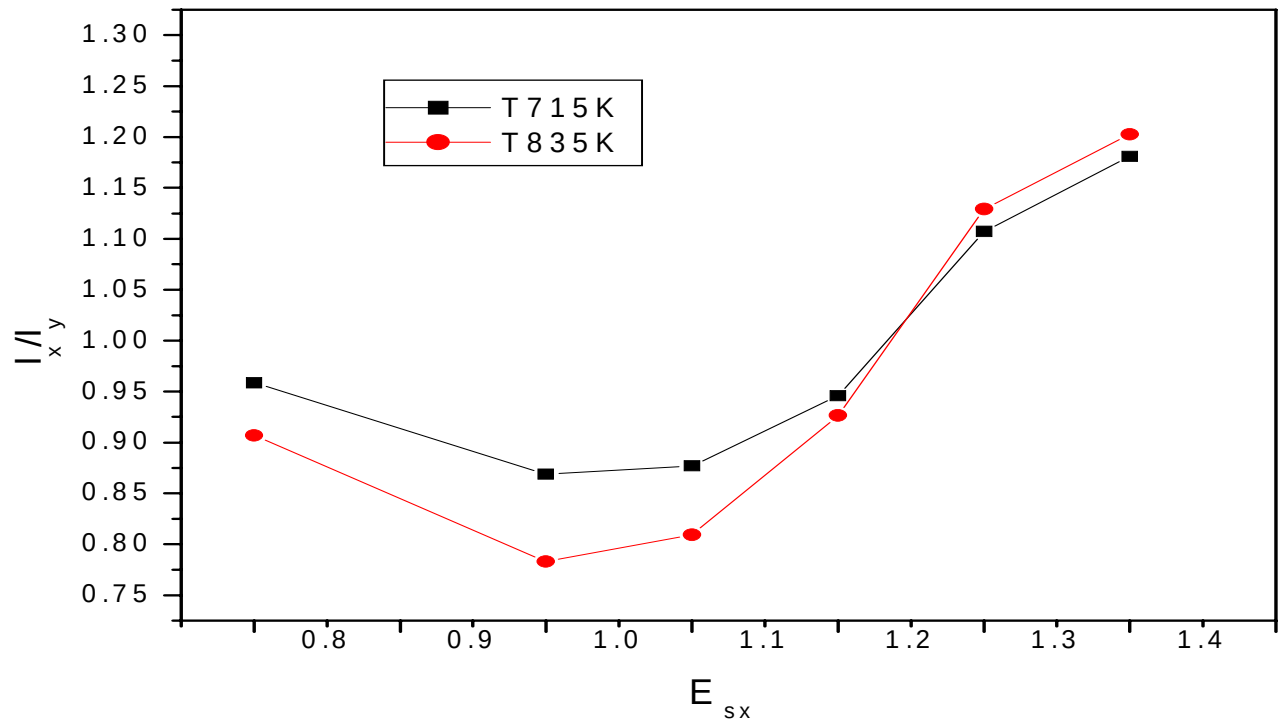

Fig.2. Variation of aspect ratio with in-channel terrace diffusion energy at temperatures T=715K and T=835K for fixed flux=0.1 ML/s, Coverage=0.075 ML and $E_{sy}$=1.3 eV

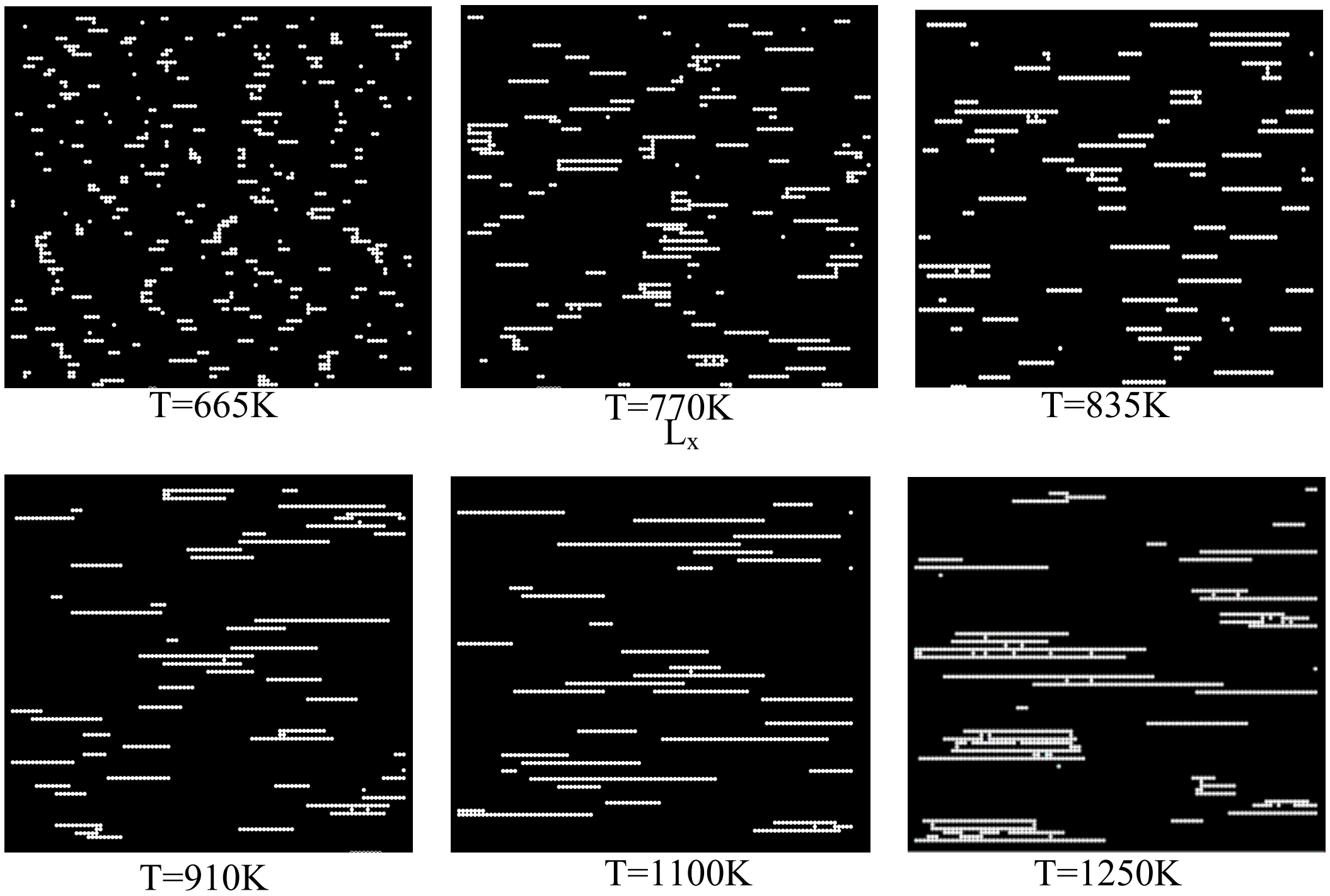

Fig.3 Temperature dependence of island morphology for model-III with $E_{sx}$=$E_{sy}$=1.3 eV, $E_{NX}$=1 eV and $E_{NY}$=0.2 eV for constant flux F=0.1ML/s for coverage θ=0.075 ML

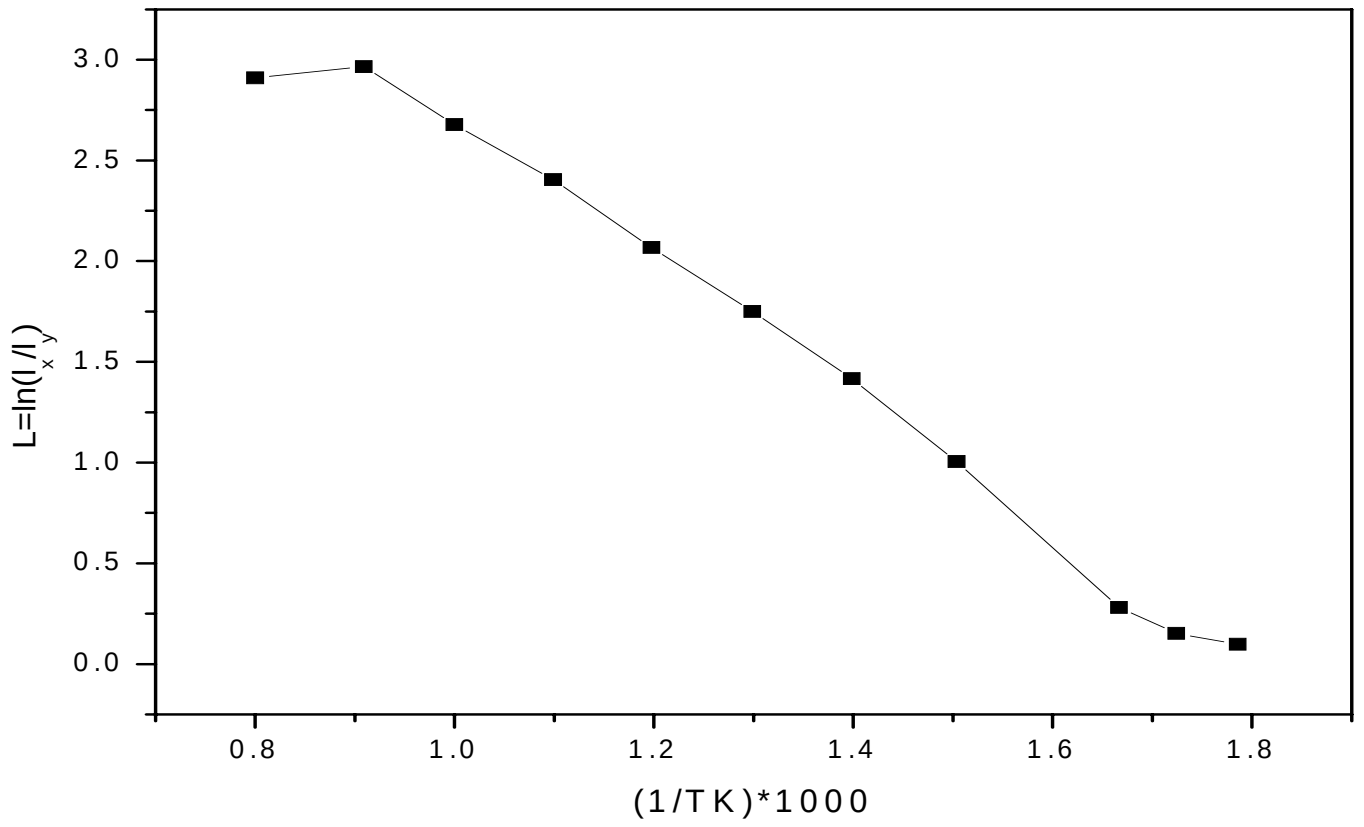

Fig.4 The aspect ratio versus temperature for model-II with $E_{sx}$=$E_{sy}$=1.3 eV, $E_{NX}$=1 eV and $E_{NY}$=0.2 eV for constant flux F=0.1ML/s for coverage θ=0.075 ML

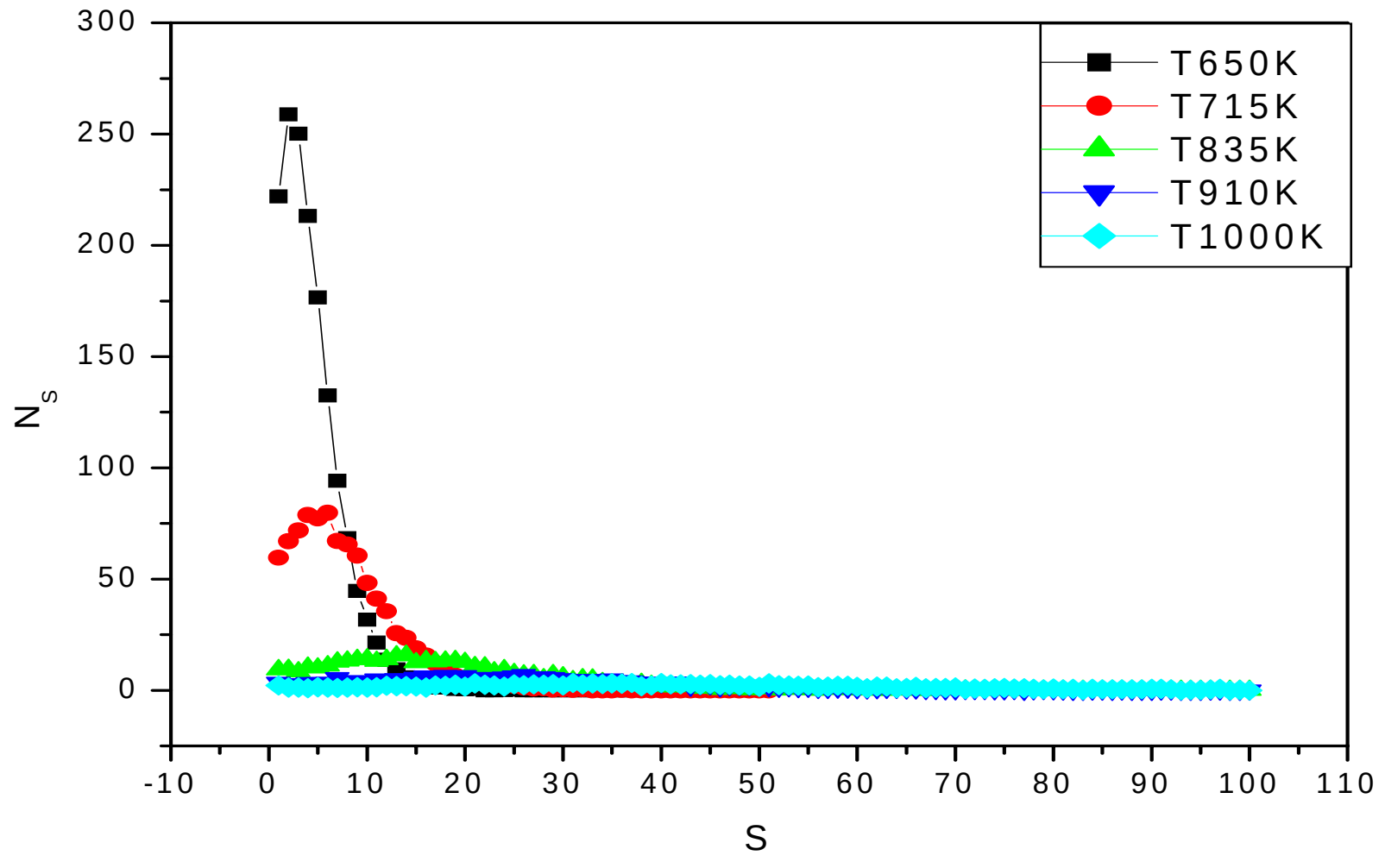

Fig 5 Island size distribution ($N_s$) versus island size (S) as a function of substrate temperature for model I, $E_{sx}$=$E_{sy}$= 1.3 eV, F = 0.1 ML/s, Cov = 0.075 ML.

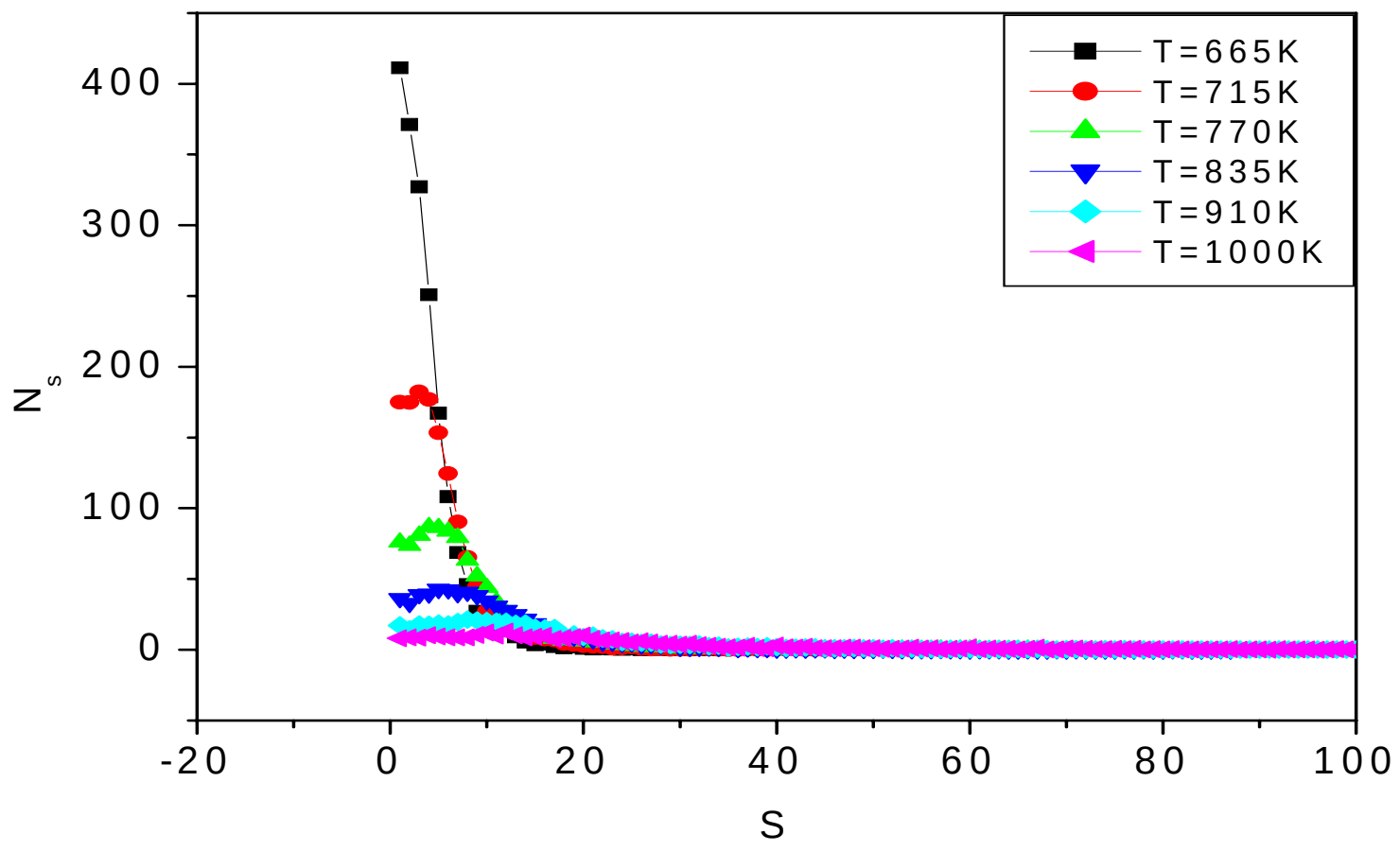

Fig. 6 Island size distribution ($N_s$) versus island size (S) for model-II at different temperature with $E_{sx}$=$E_{sy}$=1.3 eV, $E_{NX}$=1 eV and $E_{NY}$=0.2 eV for constant flux F=0.1ML/s for coverage θ=0.075 ML

also follow the similar trend as seen along the vertical panel at different energy barriers. The observed increase in island size for substrate with higher degree of anisotropy can be explained as follows. As compared to isotropic substrate, with energy barrier $E_{S,X}$ = $E_{S,Y}$=1.3 eV, in anisotropic substrate with energy barrier $E_{S,Y}$ = 1.3 and $E_{S,X}$ = 0.8 the diffusion barrier along X direction is low. Therefore, the adatoms makes larger number of jumps before another adatoms gets deposited. Due to this, nucleation probability is decreased and growth probability is increased. Therefore, islands size increases and its number density decreases. The result for effect of temperature can be also explained in similar way. When substrate temperature is high, the adatoms diffuse for longer distance within lattice and hence they encounter frequently with the existing islands and aggregate forming the island of larger size and smaller number density. We have found that the variation of N with D/F ratio shows power law behavior (not shown here). The growth exponent $\chi$ estimated from the slope were found to be $\chi$ = 0.30 and $\chi$ = 0.26 for the isotropic ($E_{S,X}$ = $E_{S,Y}$ = 1.3 eV) and anisotropic substrates respectively.

We didn't find any significant role of terrace diffusion anisotropy on island elongation at any temperature in our study. Fig.2. depicts typical plots of aspect ratio ($l_x/l_y$) with in-channel diffusion energy $E_{sx}$ at two different temperatures T = 715 K and 835 K. The plot depicts an important quantitative result that the island aspect ratio never exceeds 1.5. However, the islands are little elongated along the direction of low surface mobility. This result is similar to the results obtained by Heyn [19]. The theoretical value of aspect ratio for isotropic diffusion case is one. Our result nearly matches this value with in 10% error.

In second series of simulations (model-II), the influence of the anisotropy of the binding energies $E_{N,X}$ and $E_{N,Y}$ between atoms and island borders on island elongation i.e. aspect ratio is examined. As a convention, in-channel bonding energy ($E_{N,X}=1$ *eV)* is chosen greater than cross channel bonding energy ($E_{N,Y}=0.2$ *eV)* to make the bonds anisotropic. Due to its minor importance on island elongation and in order to reduce the number of free parameters, $E_{sx} = E_{sy} = 1.3$ eV are now chosen isotropic. The island morphology, the aspect ratio and the growth exponent as the functions of temperature have been studied in this model.

The simulation results are interpreted in terms of the anisotropic rates at which atoms detach from island borders oriented along the *X* and *Y* direction (denoted as *X* and *Y* borders). These rates are associated with the hopping frequencies of atoms with one nearest-neighbor bond. The detachment of more strongly bound atoms along X direction (i.e. at the *Y* borders) depend upon $E_S+E_{N,X}$ since bonds along the *X* - direction must be broken. Similarly the detachment of weakly bound atoms depend upon $E_S+E_{N,Y}$ since bonds along the *Y* - direction must be broken

Fig.3 and 4 show the results from a series of simulation runs where the temperature dependence of the aspect ratio is studied for fixed values of the binding energy barriers and of the flux. Three different regimes are seen in the simulations that are separated by two critical temperatures $T_1 \approx 600$K and $T_2 \approx 1100$K. In the low-temperature regime at $T<T_1$, detachment processes of atoms from islands are completely frozen out, and island growth is controlled only by adatom attachment. As a consequence, island shape is isotropic and $L \approx 0$, where $L = \ln(l_x/l_y)$ . In contrast to this, at intermediate temperatures $T_1<T < T_2$, the aspect ratio increases with T. Islands make a transition to a clearly one dimensional structure with their major axis oriented along the in-channel direction. Detachment processes of weakly bound atoms from *X* borders with rate $D_{dety}$ become apparent in this regime and the aspect ratio depends on the sum $E_S + E_{N,Y}$. We found that the mean aspect ratio mean value increases with the temperature from $L \approx 0$ to a peak value ($L = 2.9$) in this regime. This result can be explained as follows. Due to the high value of the binding energy $E_{N,X}$, the detachment rate from *Y* borders is negligibly small. In this regime, the cross-channel bonds (along Y-directions) are not stable where as in-channel bonds (along X-directions) are not broken yet. This causes the growth of islands which are oriented along the in-channel direction, whose aspect ratio increases with temperature due to the increase of the in-channel diffusion length.

Increase of temperature above the second critical temperature $T_2$ yields other qualitative change in the situation. At such high temperatures, the aspect ratio decreases with increase in temperature. In this regime, the value of $E_{N,X}$ becomes important to control the value of aspect ratio. This reflects significant detachment of strongly bond atoms from the Y boarders with rate $D_{detx}$. In this regime, very few large islands grow since diffusion is quite fast in both directions and atoms can move on the surface until they find an island where they can stick. Moreover, in-channel bond breaking begins to come into play in this regime. This causes the growth of islands with a considerable width along the cross- channel direction, and destabilizes chains of monoatomic width. Similar results were obtained by Heyn [19] and Giorgi et. al.[21].

We have studied the island size distribution $N_s$ for model-II for the different values of substrate temperature T = 650K, 715K, 770K, 835K, 910K, and 1000K. From the figure 7, it is observed that the island size distribution is drastically influenced by the variation of the substrate temperature as observed in former models. It is observed that as the value of the substrate temperature increases the position of the peak of the island size distribution curve shifts towards higher size while the peak height of the curve decreases drastically. The width of the distribution curve is observed to broaden with increase in the temperature. As studied earlier, from the comparative study of the distribution curves at the different temperatures, it is observed that the distribution curve for the higher temperature is much more scattered as compared to low temperature case. The reason for this is that at higher temperature the size of the island is bigger but the number density is much less. Due to this, the distribution curve appeared to be scattered. Comparing the curves in Fig. 6 for model II with those in Fig. 5 for model-I, it is seen that the island density is higher in model-II than in model-I where as island size is less.

**Summary and Conclusion**

From the study we found that the morphology of islands and their size distribution and number density depends upon the substrate temperature and substrate type. Higher the substrate temperature or lessen the substrate energy barrier to make anisotropic substrate, less is the number density and greater is the island size. The aspect ratio was found to be less than 1.5 for different anisotropic substrates. Hence substrate diffusion anisotropy has no significant role on island elongation. However, bonding anisotropy plays important role in island elongation. In certain range of temperature, we have found one dimensional chain of islands.